\documentclass[aps,prd,amsmath,amssymb,showpacs,showkeys]{revtex4}
\usepackage{graphicx}
\usepackage{bm}
\def\journal#1#2#3#4{{#1} {\bf #2}, #3 (#4)}

\newcommand{\be}{\begin{equation}}
\newcommand{\ee}{\end{equation}}
\newcommand{\bea}{\begin{eqnarray}}
\newcommand{\eea}{\end{eqnarray}}
\def\eq#1{(\ref{#1})}

\def\mr#1{{\mathrm{#1}}}

\begin{document}
\title{Turning points of massive particles in Schwarzschild geometry}
\author{J. Polonyi$^a$, A. Radosz$^b$, A. Siwek$^{ab}$, and K. Ostasiewicz$^b$}
\affiliation{$^a$University of Strasbourg,
High Energy Physics Theory Group, CNRS-IPHC,
23 rue du Loess, BP28 67037 Strasbourg Cedex 2, France}
\affiliation{$^b$Wroclaw University of Technology, Wroclaw, Poland}

\begin{abstract}
The stable geodesics in Schwarzschild geometry can not approach the center closer than
the radius of the photon sphere, 3/2 times the Schwarzschild radius. In other words, 
massive particles moving along geodesics that cross the photon sphere do not escape, 
they fall into the black hole.
\end{abstract}
\date{\today}

\maketitle

It is well known that there is no return for objects once they cross the horizon of the 
Schwarzschild geometry, namely when $r<r_s=2GM/c^2$. But it is less known that there is
a 50\% larger critical distance, the radius of the photon sphere, \cite{photrad} (see also \cite{twin}),
corresponding to the circular orbits of massless particles. It is shown below that already
the photon sphere radius is a point of no return for massive particles' geodesics.

\begin{figure}[h]
\begin{picture}(200,120)
\includegraphics[scale=0.6,clip=]{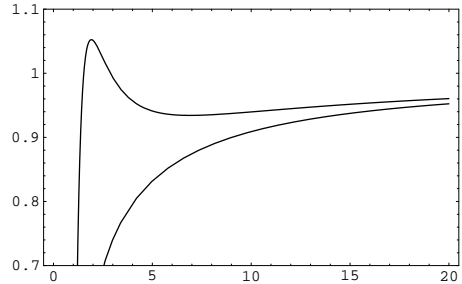}
\end{picture}
\caption{The effective potential of Eq. \eq{effpot} as the function of the dimensionless
ratio $r/r_s$ for $J/r_s=2.1$ (upper curve) and $J/r_s=1$ (lower curve).\label{effpotf}}
\end{figure}

The radial component of the velocity of a geodesics in Schwarzschild geometry takes the following form
\be
(u^r)^2=\epsilon-2V_\mr{eff}(r)
\ee
where we introduced an effective potential
\be\label{effpot}
V_\mr{eff}(r)=\frac{g_{tt}}2\left(\frac{J^2}{r^2}+1\right)
\ee
per unit mass, $g_{tt}=1-r_s/r$ and the integration constants $\epsilon$ and $J$ 
arise due to the conservation of energy and angular momentum, respectively. 
This is a monotonically increasing function of $r$ for small angular momentum, $J^2<3r_s^2$, 
but it develops local minimum $V(r_+)$ and local maximum $V(r_-)$ at
\be
r_\pm=\frac{J^2}{r_s}\left(1\pm\sqrt{1-\frac{3r_s^2}{J^2}}\right)
\ee
for larger values of angular momentum, $J^2>3r_s^2$, cf. Fig \ref{effpotf}. 
The positions of local minimum, 
$r_+$, and local maximum, $r_-$, are found to be monotonic 
functions of $J^2$, increasing and decreasing, respectively and we have 
$r_+\to\infty$ and $r_-\to r^\infty_-=3r_s/2$ as $J\to\infty$. 

The inner turning point on the geodesics, $r_\mr{min}$, satisfying the equation
\be
\epsilon-2V_\mr{eff}(r_\mr{min})=0
\ee
must be on the decreasing part of the effective potential, considered as the function of the 
radius. Therefore, $r_\mr{min}$ can not be arranged closer to the center than 
$r_\mr{min}=r^\infty_-$ i.e. within the photon sphere. One can say that the photon sphere 
is impenetrable for massive geodesics: once the freely falling massive particle crosses the 
photon sphere, it will not escape that region anymore 
(in the case of a black hole, it will inevitably reach the Schwarzschild radius).

\end{document}